\documentclass{article}

\usepackage{arxiv}

\usepackage{xcolor}

\usepackage{amsthm}

\newtheoremstyle{named}{}{}{\itshape}{}{\bfseries}{.}{.5em}{#1 \thmnote{#3}}
\theoremstyle{named}

\usepackage[utf8]{inputenc} 
\usepackage[T1]{fontenc}    
\usepackage{hyperref}       
\usepackage{url}            
\usepackage{booktabs}       
\usepackage{amsfonts}       
\usepackage{nicefrac}       
\usepackage{microtype}      
\usepackage{lipsum,cite}
\usepackage{graphicx,xcolor,amsmath,amssymb}

\graphicspath{{figures/}}
\usepackage{float}

\title{Using machine learning algorithms to determine the emotional disadaptation of a person by his rhythmogram}

\iftrue

\author{
 Sergey Stasenko\\
  Institute of Applied Physics of RAS,Russia\\
  Lobachevsky University, Russia\\
  \texttt{stasenko@neuro.nnov.ru} 
 \And
  Olga Shemagina\\
  Institute of Applied Physics of RAS,Russia\\
  \texttt{olgashemagina@gmail.com} \\
   \And   
 Eremin Evgeny\\
  Lobachevsky University, Russia\\
  \texttt{eugenevc@gmail.com} 
   \And
 Vladimir Yakhno\\
  Institute of Applied Physics of RAS, Russia\\
  \texttt{yakhno@appl.sci-nnov.ru} 
   \And
 Sergey Parin\\
 Lobachevsky University, Russia\\
  \texttt{parins@mail.ru} 
 \\
   \And
 Sofia Polevaya\\
 Lobachevsky University, Russia\\
  \texttt{sofia.polevaia@fsn.unn.ru} \\
}

\fi

\begin{document}
\maketitle

\begin{abstract}
In this study we applyed machine-learning algorithms to determine the emotional disadaptation of a person by his rhythmogram. We used the method of determining a subject level of emotional disadaptation and recording of cardiorhythmography. We show that electrocardiogram (ECG) signals can be used for the registration of the emotional disadaptation of a person.
\end{abstract}


\section{Introduction}

The development of new methods and approaches to the rapid diagnosis of stress is an urgent task, taking into account the current epidemiological (Covid-19) situation \cite{b1}. Psychological stress plays a key role in the development of many physical and neurological diseases. The term "stress" is usually used to denote both a strong adverse physical and / or psychogenic external environmental impact, and for a state of psychophysiological stress that develops under their influence, initially serving to adapt a person to new environmental conditions. Stress, as a chronic psychophysiological overstrain, can provoke the manifestation or exacerbation of symptoms of the disease, serve as one of the risk factors or aggravate the severity of the disease. Emotional overstrain reduces the productivity and quality of work performed by a person. The clinical correlates of chronic emotional stress border on neuropsychic anxiety and depressive disorders, which drastically reduce the quality of life of people \cite{b2}. Chronic emotional stress primarily negatively affects health, and indirectly causes adverse endocrine, neuromuscular, and autonomic changes \cite{b3}. Daily mental stress is the cause of many widespread serious illnesses, including hypertension, strokes, heart attacks, cancer, and more. And if the cause of acute mental stress is primarily associated with unexpected negative external influences and life changes, then the development of chronic stress is largely determined by the personal characteristics of a person and the insufficient functioning of his mechanisms of psychological adaptation. The first step to overcoming chronic emotional stress is a person's awareness of the fact that he is in a state of mental overstrain. Emotions are subjective, and the diagnosis depends on the person's ability to adequately understand and express them in words. Meanwhile, the ability to understand emotions and express them in words is not developed to the same degree in all people \cite{b4}. Therefore, the system of rapid testing of emotional disadaptation in everyday life is important for the early diagnosis of emotional overstrain and emotional exhaustion. Information about the current state of biological data allows you to provide a person with timely feedback showing the level of his mental stress, with which he can temporarily reduce this stress by switching to physical activity or other activities. Particular attention is paid to such feedback for people who are poorly aware of the peculiarities of their emotional state.

\section{Methods}

\begin{figure}[H]
   \centering
  \includegraphics[width=0.55\textwidth]{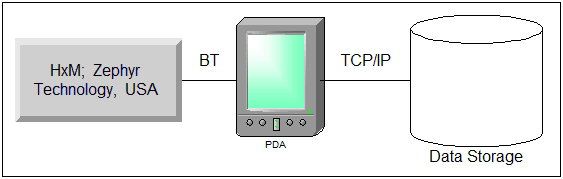}
  \caption{Scheme of the system for registering the level of emotional disadaptation according to the cardiorhythmogram}
  \label{fig1}
\end{figure}

The task of diagnosing the psychophysiological state of a person and the level of emotional disadaptation based on the physiological data of the user is solved by the method of registering emotional disadaptation according to the cardiorhythmogram, which includes the use of a mobile ECG sensor, the data from which is transmitted to a mobile application that uses a neural network algorithm previously trained at the calibration stage for classifying cardiorhythmogram data by the level of disadaptation in automatic mode 
(see Figure~\ref{fig1}). In the developed system, markers of emotional disadaptation are universal and do not require calibration for a specific user, provided that the algorithm for classifying RR intervals by labels of the level of disadaptation is pre-trained based on a database containing a sufficient number of unique user records.

At the calibration stage, a database is formed containing sequences of RR intervals of unique users extracted from ECG data, recorded in parallel with the passage of a questionnaire containing four groups of verbal characteristics that describe emotions in accordance with valence (positive/negative) and activity level (tension/relaxation) in relation to four basic personal needs: a) security; b) in independence; c) in achievement; d) in unity; (Patent RF RU2291720C1).

\subsection{Methodology for measuring the level of emotional disadaptation}

\begin{figure}[H]
   \centering
  \includegraphics[width=0.45\textwidth]{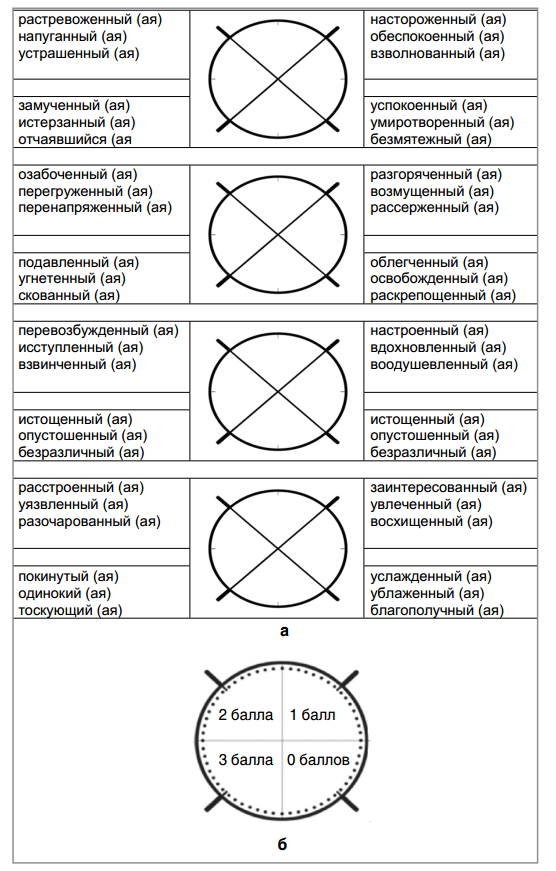}
  \caption{Methodology for determining the level of emotional disadaptation: a - type of test form; b - a circular scale for assessing the level of satisfaction for basic needs.}
  \label{fig2}
\end{figure}

To assess the level of emotional disadaptation, the test participant is asked to indicate the zone of his current state in the circular state space (see Figure~\ref{fig2}a). The boundaries of space are defined at four points of intersection of the diagonals with the circle. The boundaries are sets of synonymous adjectives that describe emotions in accordance with the modality (positive/negative) and the level of activity (tension/relaxation) in relation to four basic personal needs: a) security; b) in independence; c) in achievement; d) in unity (proximity). Depending on the position of the specified zone, the number of points scored by a person for each need is determined (see Figure~\ref{fig2}b). According to the average score, the degree of emotional disadaptation is judged as follows:
\begin{itemize}
\item 0 points - lack of emotional disadaptation (physiological relaxation);
\item 1 point - poorly expressed emotional disadaptation (physiological stress);
\item 2 - moderately expressed emotional disadaptation (pathological stress);
\item 3 - pronounced emotional disadaptation (pathological relaxation).
\end{itemize}

Telemetric measurements of the electrocardiogram were carried out using a software and hardware complex consisting of a miniature wireless ECG sensor (HxM; Zephyr Technology, USA) and a smartphone with specialized software (see Figure~\ref{fig1});
Based on the collected data for different users or one user, the algorithm for classifying the sequence of RR intervals was trained according to the labels of the level of emotional disadaptation of the user;
The pre-trained classification algorithm is applied to the new data of the sequence of RR intervals to determine the level of disadaptation in automatic mode without using a questionnaire.

\subsection{Data for building a classifier}

\begin{figure}[H]
   \centering
  \includegraphics[width=0.65\textwidth]{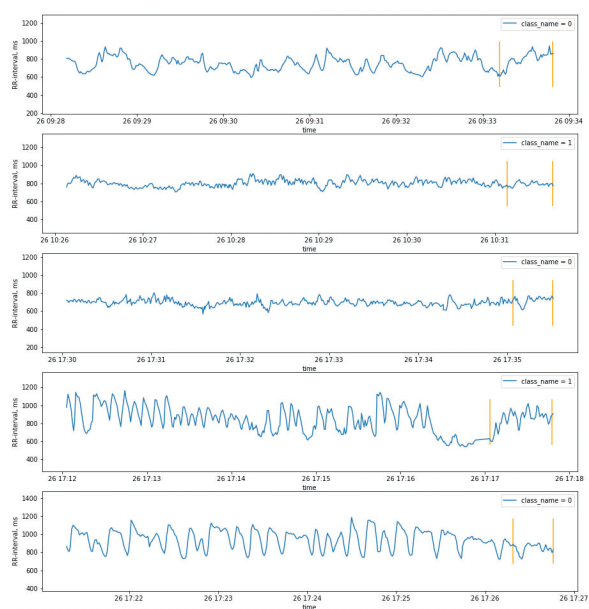}
  \caption{Examples of cardiograms. Yellow marks indicate the interval for passing the test. Belonging to a class ($class\_name$) is marked on the graph.}
  \label{fig3}
\end{figure}

To build a classifier, data from users' heart rhythmograms were used (see Figure~\ref{fig3}). Each entry corresponded to the level of emotional disadaptation, determined using a questionnaire. The collected database consists of 2222 unique records.

Each entry is a set of files with the following content:

\begin{itemize}
\item rr\underline{\hspace{.10in}}filter.csv - dependence of the RR-interval ($RR\_filter$) value on time (ms) within 5 minutes before passing the test to determine the level of emotional disadaptation plus time of passing the test.
\item info.csv - recording start time ($time$), post ID ($session\_id$), subject identifier ($person\_id$), age ($old$) and subject's gender ($gender$)
\item uad.csv - time stamps of the beginning ($ms\_begin$) and end ($ms\_end$) of the test (in milliseconds from the start time of the recording), the result of the test - different degrees of satisfaction of the following four basic personal needs: safety, independence, achievement, unity - closeness (U1; U2; U3; U4), the subjective level of emotional disadaptation, the average value of the degree of satisfaction of four basic personal needs (medium).
\end{itemize}

The time for passing the test for all subjects was different, and, accordingly, the duration of the record varied, which is inconvenient from the point of view of the algorithms used in the construction of the classifier, so all records were divided into parts of 100 samples, starting from the end of the record, since the fragments of the record corresponding to at the time of filling out the questionnaire, seemed to us the most informative. The rest of the record containing less than 100 samples is discarded. All data were standardized and divided into 2 classes: records with an emotional disadaptation level of 0 or 1 according to the results of the questionnaire belonged to class 0, corresponding to the absence of disadaptation, the rest of the records belonged to class 1 and corresponded to the presence of emotional disadaptation.

\subsection{Machine-learning algorithms}

The following algorithms were used to build the classifiers:
\begin{enumerate}
\item Logistic regression \cite{b8} - is a method for constructing a classifier based on linear models of the following type: $\hat{y}(w, x) = w_0 + w_1 x_1 + ... + w_p x_p$, where $\hat{y}(w, x)$ is the predicted value. The task of training a linear classifier is to adjust the weight vector $w$ based on the sample $X^m$. In logistic regression, for this, the problem of empirical risk minimization is solved with a special type of loss function: $\min_{w, c} \frac{1}{2}w^T w + C \sum_{i=1}^n \log(\exp(- y_i (X_i^T w + c)) + 1)$. To solve the optimization problem the Broyden — Fletcher — Goldfarb — Shanno algorithm is used.
\item Easy ensamble \cite{b9} - the idea is to train an ensamble of classifiers, each of which makes its own decision about the belonging of an object to a certain class.  The final decision is made by the result of voting.  The algorithm is used for unbalanced data.  The entire minority class and a subset of the majority class are used to train classifiers. Main parameters of the whole ensamble are a number of learners (we use 10) and an algorythm to learn each classifier in ensamble (we use AdaBoost).
\item Gradient boosting \cite{b10} - the idea is to combine several weak classifiers based on decision trees into one strong classifier.
\end{enumerate}
These algorythms are implemented in Scikit-learn library \cite{b11} and Imbalanced-learn library \cite{b12}.

\section{Results}

80\% of the generated database was used to train classifiers. The remaining 20\% were the test sample. In the process of testing, the record, for which it was necessary to make a decision about belonging to a particular class, was divided into parts of 100 samples, for each part the classifier made a decision, and the final decision was formed as the arithmetic mean of the values of the class labels of each of the parts. Rounding was carried out according to the rules of mathematical rounding.

\begin{figure}[H]
   \centering
  \includegraphics[width=0.65\textwidth]{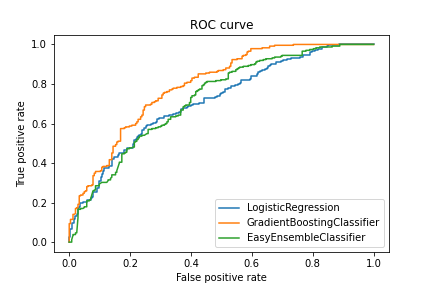}
  \caption{Classification result based on ROC-curve for three algorithms of classification: logistic regression, gradient boosting, and easyEnsembleClassifier.}
  \label{fig4}
\end{figure}

The result of the work of classifiers built on the basis of the algorithms described above is shown in (see Figure~\ref{fig4}).

\begin{figure}[H]
   \centering
  \includegraphics[width=0.65\textwidth]{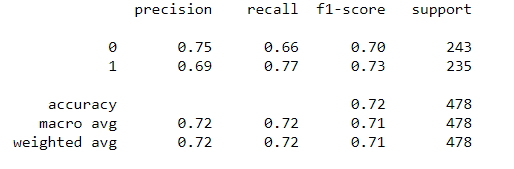}
  \caption{Precision Recall Table for Gradient Boosting Classification.}
  \label{fig5}
\end{figure}

As can be seen from the graph above, the classifier based on gradient boosting showed the best result (see Figure~\ref{fig5}).

\subsection{Using dynamic spectra to build a classifier}

\begin{figure}[H]
   \centering
  \includegraphics[width=0.75\textwidth]{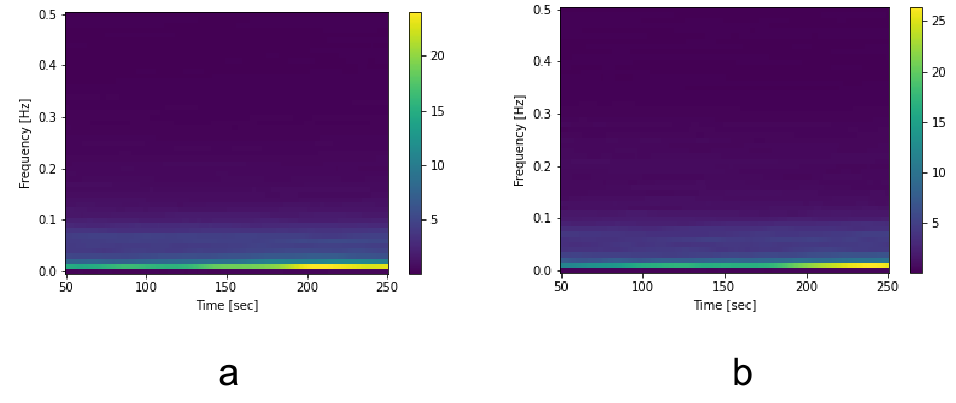}
  \caption{Averaged spectrogram for training sample records; a - belonging to class 0, b - belonging to class 1.}
  \label{fig6}
\end{figure}

\begin{figure}[H]
   \centering
  \includegraphics[width=0.55\textwidth]{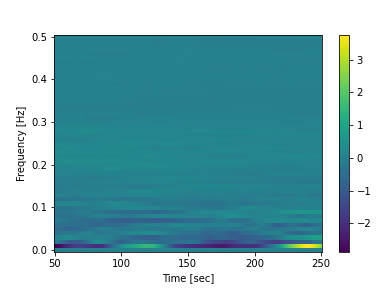}
  \caption{Difference spectrogram.}
  \label{fig7}
\end{figure}

The disadvantage of using original recordings of rhythmograms is the impossibility of their temporal synchronization, especially in the case of splitting records into fragments. Therefore, it seemed logical to move to such a feature space that would level this shortcoming.
To build a classifier based on dynamic spectra, the original signal was divided into fragments of 300 samples, the spectrum was calculated in a window of 100 samples. By shifting the window by 1 count and calculating the spectrum in a new window, we obtain a spectrogram (see Figures~\ref{fig6} and ~\ref{fig7}).

\begin{figure}[H]
   \centering
  \includegraphics[width=0.65\textwidth]{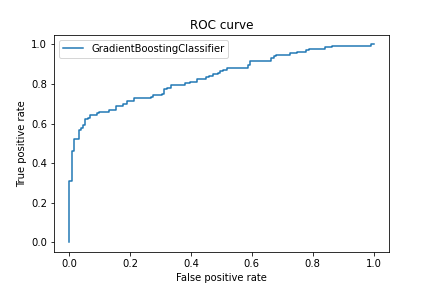}
  \caption{Classification result based on ROC-curve for the algorithm of classification built on spectrograms - gradient boosting.}
  \label{fig8}
\end{figure}

\begin{figure}[H]
   \centering
  \includegraphics[width=0.65\textwidth]{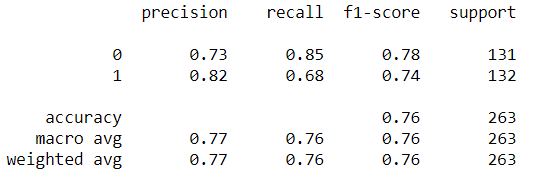}
  \caption{Precision Recall Table for Gradient Boosting Classification built on spectrograms.}
  \label{fig9}
\end{figure}

To build a classifier based on spectrograms, each spectrogram is converted into a row vector. As a classification algorithm, we choose the well-proven GradientBoostingClassifier.
The result of the work of the constructed classifier exceeds the result of the work of the classifier that uses the original rhythmograms (see Figures~\ref{fig8} and \ref{fig9}).

\section{Conclusion and Discussion}

In this paper we suggest a novel a method for determining the level of emotional disadaptation according to a person's cardiorhythmogram. The determination takes place automatically. Data from a portable heart sensor attached to a person is transmitted via Bluetooth to a mobile device, where the level of emotional disadaptation is determined using a pre-trained neural network algorithm. As a neural network algorithm, it is advisable to use a classifier trained on the basis of spectrograms, and not on the initial data of cardiorhythmograms.

\section{Acknowledgements}

The work in terms of data preprocessing was supported by the Russian Ministry of Science and Education project number 075-15-2021-634 and in term of data analisys was supported by the frames of the Governmental Project   of   the   Institute   of   Applied   Physics  RAS  (Project number  0030-2021-0014).





\end{document}